\documentclass[12pt]{article}
\usepackage{geometry}             
\usepackage{graphicx}
\usepackage{amssymb}
\usepackage{amsmath}
\usepackage{epstopdf}

\usepackage[hyperindex=true,
          pdfstartview=FitH,
          bookmarksnumbered=true,
          bookmarksopen=true,
          citecolor=blue,
          linkcolor=blue,
          colorlinks=true,
          pdfborder=001,
          unicode]{hyperref}

\begin{document}

\title{Gauss-Bonnet as effective cosmological constant}
\author{Liu Zhao, Kun Meng\\
School of Physics, Nankai university, Tianjin 300071, China\\
email: {\it lzhao@nankai.edu.cn}, {\it kunmeng@mail.nankai.edu.cn} }
\date{\today}                             % Activate to display a given date or no date
\maketitle

\begin{abstract}
It is known that Gauss-Bonnet terms in higher dimensional gravity can produce 
an effective cosmological constant. We add extra examples to this picture by presenting 
explicitly two branches of accelerating vacuum solutions to the Einstein-Gauss-Bonnet 
gravities with a bare cosmological constant in 5 and 6 dimensions. Both branches of 
solutions are of constant curvature and the effective cosmological constants are 
independent of the acceleration parameter. One branch 
(the ``$-$'' branch)  of the solutions is well defined in the limit when the Gauss-Bonnet 
parameter approaches zero, in which case the effective cosmological constant becomes 
identical with the bare value, while the other (i.e.  the ``$+$'') branch is singular in the 
same limit, and beyond this singular limit, the effective cosmological constant is inversely 
proportional to the Gauss-Bonnet parameter with a negative constant of proportionality 
when the bare value vanishes.
\end{abstract}

\section{Introduction}

Gravitational theories with higher curvature terms have attracted 
considerable attention. Among these, the Lovelock family \cite{Lovelock:1971p5956} of 
gravity theories -- of which Einstein-Gauss-Bonnet (EGB)
gravity is a member involving contributions in curvature 
tensor up to second order  -- are of particular interests because these are the only 
theories which degenerate 
naturally into Einstein gravity in 4-dimensions \cite{Lovelock:1972p5949}. Exact  static  
spherically  symmetric
black  hole  solutions  of  EGB gravity have been found
in Refs. \cite{Boulware:1985p5985, Wheeler},  and of the 
Einstein-Maxwell-Gauss-Bonnet (EMGB) and 
Born-Infeld-Gauss-Bonnet models
in Refs. \cite{Wiltshire, Neupane:2002p6109}.
Black hole solutions with nontrivial topology have also been studied in
\cite{Cai:1998p6110, Cai:2001p6111, Aros:2000p6112, Cho:2002p6113, 
Dehghani:2004p869}.
Other solutions like charged rotating black brane
solutions \cite{Dehghani:2002p6114}, Taub-NUT solutions in 
EMGB gravity in $2k + 2$
dimensions \cite{Dehghani:2006p2309}, nonstatic brane cosmology
solution \cite{Charmousis:2002p6115} and numeric black hole
solutions \cite{Brihaye:2011p6116} have also been studied. The
thermodynamics of various black holes have also been studied in
Refs. \cite{Wiltshire, Neupane:2002p6109, Myers:1988p5951, Myers2, 
Cvetic:2001p6166, Nojiri:2001p6167, Cho:2002p5953}.

In this paper, we shall consider another aspect of 
EGB gravity, i.e. the effect of Gauss-Bonnet term on producing effective 
cosmological constant. That the Gauss-Bonnet term can produce effective cosmological 
constant has been known for a long time, see, e.g. \cite{Boulware:1985p5985}. In 
\cite{Cai:2001p6111} and \cite{Dehghani:2004p869}, the same effect were also 
reproduced. Unlike the solutions 
given in \cite{Boulware:1985p5985} \cite{Cai:2001p6111} \cite{Dehghani:2004p869}, 
the explicit solutions we shall present is written in a special form which is conformal to 
the standard (A)dS metrics and the conformal factor depends on an angular coordinate, 
thus making the spacetime anisotropic. Similar anisotropic vacuum solutions in Einstein 
gravity were studied in  \cite{Zhao:2011p693} and are known to represent accelerating 
vacua. We believe that such vacuum solutions 
for EGB gravities were not discussed before in the literatures.

\section{Einstein vacua for Gauss-Bonnet gravity in 5-dimensions}

We begin by writing down the action of EGB gravity in $n$-dimensions:
\begin{align*}
&I=\frac{1}{16\pi G}\int \mathrm{d}^n x
\sqrt{-g}\left[R-2\Lambda+\xi \mathcal{L}_{\mathrm{GB}}\right],\\
&\mathcal{L}_{\mathrm{GB}}=R_{\mu\nu\gamma\delta}R^{\mu\nu\gamma
\delta}-4R_{\mu\nu}R^{\mu\nu}+R^2, 
\end{align*}
where $G$ is the $n$-dimensional Newton constant, $\Lambda$ is the bare 
cosmological constant and $\xi$ is the Gauss-Bonnet parameter.

The field equation that follow from the action reads
\begin{align}
R_{\mu\nu}-\frac{1}{2}g_{\mu\nu}R + \Lambda g_{\mu\nu} + \xi H_{\mu\nu}
=0, \label{eqm1}
\end{align}
where
\begin{align}
H_{\mu\nu}=2\left(
R_{\mu\lambda\rho\sigma}R_{\nu}{}^{\lambda\rho\sigma}
-2R_{\mu\rho\nu\sigma}R^{\rho\sigma}-2R_{\mu\sigma}R_{\nu}{}^{\sigma}
+RR_{\mu\nu}\right)
-\frac{1}{2}\mathcal{L}_{\mathrm{GB}} g_{\mu\nu}. \label{eqm2}
\end{align}

In this section, we shall consider the case $n=5$. Using the metric ansatz 
\begin{align}
  \mathrm{d} s^2 = \frac{1}{\alpha^2(x+y)^2}\left[-Y(y)\mathrm{d} t^2+
  \frac{\mathrm{d} y^2}{Y(y)}
  +\frac{\mathrm{d} x^2}{X(x)}+X(x)\left(\frac{\mathrm{d} z^2}{Z(z)}+Z(z)
  \mathrm{d} \phi^2 \right)\right] \label{ansatz}
\end{align}
and by use of some computer algebra, we found that the following are exact solutions to 
the equations of motion (\ref{eqm1})-(\ref{eqm2}):
\begin{align*}
  X(x)&=\frac{1}{2}C_1x^2+C_2x+C_3,\\
  Y(y)&=-\left(\frac{1}{2}C_{1}y^{2}-C_{2}y+C_{3}\right)
  +\frac{1}{12\alpha^2\xi}\left(3+\delta\sqrt{12\xi\Lambda+9}\,\right),\\
  Z(z)&=\frac{1}{2}\left(C_1C_3-\frac{1}{2}C_2^2\right)z^2+C_4z+C_5,
\end{align*}
where $\delta=\pm1$, $C_{1}\sim C_{5}$ are integration constants and $\alpha$ is a 
free parameter introduced in the metric ansatz. We can freely take $C_{2}=C_{4}=0$ 
and $\mathrm{abs}(C_{3})=1$  by use of the coordinate shift degrees of freedom. 
Further more, by use of the coordinate rescaling degrees of freedom, we can make
$\mathrm{abs}(C_{1})=2$. We also choose $C_{1}C_{3}=-2$ and $C_{5}=1$ for 
simplicity. Then the solutions will read
\begin{align}
  X(x)&=\epsilon(1-x^2), \label{x5}\\
  Y(y)&=\epsilon(y^{2}-1)
  +\frac{1}{12\alpha^2\xi}\left(3+\delta\sqrt{12\xi\Lambda+9}\,\right),
  \label{y5}\\
  Z(z)&=1-z^{2}, \label{z5}
\end{align}
where $\epsilon=\pm1$ and is independent of $\delta$. 

The above vacuum solutions were written in an unfamiliar coordinate system. To make 
them more recognizable, we need to make some coordinate transformations. 
For all values of $\epsilon$, we change $y$ and $z$  as follows,
\begin{align*}
& t\rightarrow \tau=\frac{t}{\alpha},\\
&y\rightarrow r=\frac{1}{\alpha y},\\
&z\rightarrow \theta_{2}=\arccos(z).
\end{align*}
The $x$ coordinate transformation will depend on the value of $\epsilon$. For $
\epsilon=1$, we take the transformation
\begin{align*}
&x\rightarrow \theta_{1}=\arccos(-x),
\end{align*}
and for the case $\epsilon=-1$, we take  
\begin{align*}
&x\rightarrow \theta_{1}=\mathrm{arccosh}(-x).
\end{align*}
In all cases the coordinates $t$ and $\phi$ are kept unchanged. The resulting metrics 
after the above transformations are given as follows. For $\epsilon=+1$, we have
\begin{align}
ds^{2}&=\frac{1}{(1- \alpha r \cos\theta_{1})^{2}}\left(
-f_{1}(r)d\tau^{2}+ \frac{1}{f_{1}(r)}dr^{2} + r^{2} d\Omega_{3+}^{2}\right),
\label{m1}\\
f_{1}(r)&= 1-\frac{\kappa_{1} r^{2}}{\ell_{1}^{2}}\equiv 1-r^{2}\left(\alpha^{2}-
\frac{3+\delta\sqrt{12\xi\Lambda+9}}{12\xi}\right),
\label{f1}\\
d\Omega_{3+}^{2}&=d\theta_{1}^{2}+\sin^{2}\theta_{1}(d\theta_{2}^{2}+
\sin^{2}\theta_{2}d\phi^{2}),
\end{align}
where $\kappa_{1}=\pm1$ depending on the values of $\delta, \xi, \Lambda$ and $
\alpha$. Similarly, for $\epsilon=-1$, the corresponding metric becomes
\begin{align}
ds^{2}&=\frac{1}{(\alpha r \cosh\theta_{1}-1)^{2}}\left(
-f_{2}(r)d\tau^{2}+ \frac{1}{f_{2}(r)}dr^{2} + r^{2} d\Omega_{3-}^{2}\right),
\label{m2}\\
f_{2}(r)&=-1 +\frac{\kappa_{2}r^{2}}{\ell_{2}^{2}}\equiv -1+r^{2}\left(\alpha^{2}+
\frac{3+\delta\sqrt{12\xi\Lambda+9}}{12\xi}\right),
\label{f2}\\
d\Omega_{3-}^{2}&=d\theta_{1}^{2}+\sinh^{2}\theta_{1}(d\theta_{2}^{2}+
\sin^{2}\theta_{2}d\phi^{2}),
\end{align}
where, again, $\kappa_{2}=\pm1$ depending on the values of $\delta, \xi, \Lambda$ 
and $\alpha$. 

The metric (\ref{m1}) is explicitly conformal to the standard (A)dS 
metric. Without much effort, it can be shown that, although we obtained the metric 
(\ref{m1}) as an exact solution to the equations of motion (\ref{eqm1}) (with the 
insertion of (\ref{eqm2})) of EGB gravity, it is indeed also a solution to the 
pure Einstein equation
\begin{align}
R_{\mu\nu}-\frac{1}{2}g_{\mu\nu}R +\Lambda_{\mathrm{eff}}g_{\mu\nu}=0
\label{eins}
\end{align}
with the effective cosmological constant $\Lambda_{\mathrm{eff}}$ given as
\begin{align}
\Lambda_{\mathrm{eff}}=6\left(\frac{\kappa_{1}}{\ell_{1}^{2}}-\alpha^{2}\right)
=-\frac{3+\delta\sqrt{12\xi\Lambda+9}}{2\xi}. \label{effl}
\end{align}
The form of the metric (\ref{m1}) has already been found in the previous work 
\cite{Zhao:2011p693} by one of the authors as an exact solution of Einstein gravity with 
non vanishing cosmological constant.

Apart from the overall conformal factor, the metric (\ref{m2}) is identical to the 
well known topological AdS black hole metric \cite{Brill, Emparan, Birmingham} with 
mass set to zero and curvature of spacial sections set to negative constants. The 
complete metric (\ref{m2}) also satisfies the effective Einstein equation (\ref{eins}) with 
the effective cosmological constant given by
\begin{align}
\Lambda_{\mathrm{eff}}=6\left(\alpha^{2}-\frac{\kappa_{2}}{\ell_{2}^{2}}\right)
=-\frac{3+\delta\sqrt{12\xi\Lambda+9}}{2\xi}.\label{effl2}
\end{align}
Thus both the cases $\epsilon=+1$ and $\epsilon=-1$ correspond to the same effective 
cosmological constant. 

For both the metric (\ref{m1}) or the metric (\ref{m2}), it can be easily checked that 
the parameter $\alpha$ has a very clear physical meaning. It corresponds to the 
magnitude of proper acceleration of the origin of the coordinate system being used.
It should be remarked that accelerating vacua with conformal factor as given in 
(\ref{m2}) (i.e. involving hyperbolic cosine in the denominator) have not been found 
before.

It follows from (\ref{effl})-(\ref{effl2}) that the effective cosmological constant is 
independent of the acceleration parameter. Actually, by taking the $\alpha=0$ limit in  
(\ref{m1}), the effective cosmological constant can be readily read off 
because the metric falls back to that of the standard (A)dS. This is in sharp contrast to 
the case of Einstein gravity, for which it is known that the presence of the accelerating 
parameter $\alpha$ will shift the corresponding cosmological constant 
\cite{Zhao:2011p693}.

To see how the dependence on the bare cosmological 
constant $\Lambda$, the Gauss-Bonnet parameter $\xi$ and the 
signature $\delta$ affects the value of the effective cosmological constant, it is necessary 
to take various limits. First we consider 
the ``$-$'' branch (i.e. $\delta=-1$). In this case $\Lambda_{\mathrm{eff}}$ is well 
behaved in the limit $\xi \rightarrow 0$, yielding
\[
\lim_{\xi=0,\,\delta=-1}\Lambda_{\mathrm{eff}}=\Lambda.
\]
On the other hand, the ``$+$'' branch does not have a well defined limit while  $\xi 
\rightarrow 0$. So, we may consider another limit, i.e. $\Lambda\rightarrow 0$. 
The limiting value of $\Lambda_{\mathrm{eff}}$ in this case reads
\[
\lim_{\Lambda=0}\Lambda_{\mathrm{eff}}=-\frac{3}{2\xi}(1+\delta),
\]
which is nonvanishing only for $\delta=+1$. Therefore,  in order to produce a nonzero 
effective constant starting from a vanishing bare value, we need to take $\delta=+1$, and
the corresponding effective cosmological constant is
\[
\Lambda_{\mathrm{eff}}=-\frac{3}{\xi}.
\]

\section{Effective cosmological constant in 6D}

In the last section we have seen that the presence of the Gauss-Bonnet parameter $\xi$ can produce an effective cosmological constant in 5-dimensions. It is natural to ask 
whether this is an accidental fact in 5D or is a general property in all higher dimensions. 
In this section, we shall try to partially settle this question by presenting explicit results 
in the case of 6-dimensions.

We start by mimicking the ansatz (\ref{ansatz}) by writing down the following metric:
\begin{align*}
  \mathrm{d} s^2 &= \frac{1}{\alpha^2(x+y)^2}\\
  &\quad \times
  \left\{-Y(y)\mathrm{d} t^2+\frac{\mathrm{d} y^2}{Y(y)}
  +\frac{\mathrm{d} x^2}{X(x)}+X(x)\left[\frac{\mathrm{d} z^2}{Z(z)}+Z(z)
  \left(\frac{\mathrm{d} w^2}{W(w)}+W(w)\mathrm{d} \phi^2 \right)\right]\right\}.
\end{align*}
Inserting into the equation of motion (\ref{eqm1})-(\ref{eqm2}), we get the following 
solution,
\begin{align*}
  X(x)&=\frac{1}{2}C_1x^2+C_2x+C_3,\\
  Y(y)&=-\left(\frac{1}{2}C_{1}y^{2}-C_{2}y+C_{3}\right)
  +\frac{1}{60\alpha^2\xi}\left(5+\delta\sqrt{60\xi\Lambda+25}\,\right),\\
  Z(z)&=\frac{1}{2}\left(C_1C_3-\frac{1}{2}C_2^2\right)z^2+C_4z+C_5,\\
  W(w)&= \frac{1}{2}\left(C_1C_3C_5-\frac{1}{2}C_4^2-\frac{1}{2}
  C_2^2C_5\right)w^2+C_6 w+C_7,
\end{align*}
where $\delta=\pm1$ and $C_{1}\sim C_{7}$ are integration constants. By properly 
choosing the values of these constants using various coordinate choice freedoms, we can 
simplify the solution into the form
\begin{align}
  X(x)&=\epsilon(1-x^2), \label{x6}\\
  Y(y)&=\epsilon(y^{2}-1) 
  +\frac{1}{60\alpha^2\xi}\left(5+\delta\sqrt{60\xi\Lambda+25}\,\right),\\
  Z(z)&=1-z^{2},\\
  W(w)&= 1-w^{2}. \label{w6}
\end{align}
One can proceed as in the previous section to rewrite the solution in a form which is 
conformal to the (A)dS metric in the standard form. However, omitting the details, we 
jump to the conclusion that the above metric is also a solution to the Einstein equation 
(\ref{eins}) with the effective cosmological 
constant given by
\[
\Lambda_{\mathrm{eff}}=-\frac{5+\delta\sqrt{60\xi \Lambda +25}}{6\xi}.
\]
For vanishing $\Lambda$, $\Lambda_{\mathrm{eff}}$ is nonzero only in the ``$+$'' 
branch and the corresponding value is
\[
\Lambda_{\mathrm{eff}}=-\frac{5}{3\xi}.
\]

\section{Discussion}

We have shown that EGB gravities in 5 and 6 dimensions possess accelerating 
vacua which are identical to Einstein vacua with effective cosmological constants. The 
solutions can be divided into two branches by choosing different signatures $\delta$, 
just like in the case in the absence of the acceleration parameter 
\cite{Boulware:1985p5985}. 
The ``$-$'' branch is well behaved when the Gauss-Bonnet parameter $\xi\rightarrow 
0$, in which case the effective cosmological constant approaches 
the bare value. In contrast, the ``$+$'' branch is not well behaved in the limit  $\xi
\rightarrow 0$, and for $\xi\ne 0$ the effective cosmological constant does not vanish 
even the bare value is zero. In both 5D and 6D cases the effective cosmological 
constant in the ``$+$'' branch with vanishing bare value is inversely proportional to the 
Gauss-Bonnet parameter, with the constant of proportionality being a negative number. 
Thus, for instance, if we need a small and positive effective cosmological constant, the 
Gauss-Bonnet parameter must be negative and have a very large absolute value. On the 
other hand, if the Gauss-Bonnet parameter were taken to be positive, the effective 
cosmological constant will be negative. 

It is remarkable that the effective
cosmological constants does not depends on the acceleration parameter 
(as apposed to the case of Einstein gravity), 
and the concrete values we get in 5D and 6D agree with the corresponding values
in the non accelerating cases \cite{Boulware:1985p5985}.
We speculate that these properties may continue to hold in even higher 
dimensions. It is interesting to ask whether higher order Lovelock parameters can also  
produce effective cosmological constants. We leave this question for later considerations.

In closing this article, let us pay some words toward the similar effects in pure GB gravity in the absence of Einstein term. It is not quite difficult to check that the solution
(\ref{x5})-(\ref{z5}), with (\ref{y5}) replaced by
\begin{align*}
  Y(y)&=\epsilon(y^{2}-1)
  +\frac{1}{6\alpha^2\xi}\left(\delta\sqrt{3\xi\Lambda}\,\right),
\end{align*}
satisfies the equation of motion for the so-called pure GB gravity in 5D, 
\[
\Lambda g_{\mu\nu}+\xi H_{\mu\nu}=0.
\]
Similarly, the solution (\ref{x6})-(\ref{w6}), with $Y(y)$ replaced by
\begin{align*}
  Y(y)&=\epsilon(y^{2}-1) 
  +\frac{1}{30\alpha^2\xi}\left(\delta\sqrt{15\xi\Lambda}\,\right),
\end{align*}
also satisfy the equation of motion for pure GB gravity in 6D. The corresponding solutions are again Einstein vacua, with the effective cosmological constants 
given by
\begin{align*}
\Lambda_{\mathrm{eff}}=-\frac{\delta\sqrt{3\xi\Lambda}}{\xi}
\end{align*}
in the 5D case and 
\begin{align*}
\Lambda_{\mathrm{eff}}=-\frac{\delta\sqrt{15\xi\Lambda}}{3\xi}
\end{align*}
in the 6D case. So it is clear that for pure GB gravity, the effective 
cosmological constants will approach zero in both branches of accelerating vacua
if the bare value vanishes.

\section*{Acknowledgment} 

This work is supported by the National  Natural Science Foundation of 
China (NSFC) through grant No.10875059. L.Z. would like to thank the 
organizer and participants of ``The advanced workshop on Dark Energy 
and Fundamental Theory'' supported by the Special Fund for Theoretical 
Physics from the National Natural Science Foundation of China with grant 
no: 10947203 for discussions.

%\bibliographystyle{utcaps}
%\bibliography{../../BibLibrary/papers2.bib,../../BibLibrary/papers1.bib}

\begin{thebibliography}{10}

\bibitem{Lovelock:1971p5956}
D.~Lovelock, ``The Einstein tensor and its generalizations,'' J. Math. Phys.12 (1971) 498.

\bibitem{Lovelock:1972p5949}
D.~Lovelock, ``The Four-Dimensionality of Space and the Einstein Tensor,''
  J. Math. Phys.13 (1972) 874.

\bibitem{Boulware:1985p5985}
D.~Boulware and S. Deser, ``String-generated gravity models,'' Phys. Rev.
  Lett. (1985).

\bibitem{Wheeler}J. T. Wheeler, ``Symmetric solutions to the Gauss-Bonnet extended Einstein equations,'' Nucl. Phys. B268 (1986) 737.

\bibitem{Wiltshire}D. L. Wiltshire, ``Spherically symmetric solutions of Einstein-Maxwell theory with a Gauss-Bonnet term'', Phy. Lett. B
V169, No.1 (1986) 36-40.

\bibitem{Neupane:2002p6109}
I.~P. Neupane, ``Black hole entropy in string-generated gravity models,'' Phys.Rev. D67: 061501, 2003 [arXiv:
  \href{http://www.arXiv.org/abs/hep-th/0212092v3}{{\tt hep-th/0212092}}].

\bibitem{Cai:1998p6110}
R.-G. Cai and K.-S. Soh, ``Topological black holes in the dimensionally
  continued gravity,'' Phys. Rev. D59: 044013, 1999 [arXiv: \href{http://www.arXiv.org/abs/gr-qc/9808067v2}{{\tt gr-qc/9808067}}].

\bibitem{Cai:2001p6111}
R.-G. Cai, ``Gauss-Bonnet Black Holes in AdS Spaces,'' Phys.Rev.D65: 084014, 2002 [arXiv: \href{http://www.arXiv.org/abs/hep-th/0109133v2}{{\tt
  hep-th/0109133}}].

\bibitem{Aros:2000p6112}
R.~Aros, R.~Troncoso, and J.~Zanelli, ``Black holes with topologically
  nontrivial AdS asymptotics,'' Phys. Rev. D63: 084015, 2002 [arXiv:
  \href{http://www.arXiv.org/abs/hep-th/0011097v2}{{\tt hep-th/0011097}}].

\bibitem{Cho:2002p6113}
Y.~M. Cho and I.~P. Neupane, ``Anti-de Sitter Black Holes, Thermal Phase
  Transition and Holography in Higher Curvature Gravity,'' Phys. Rev. D66: 024044, 2002 [arXiv: \href{http://www.arXiv.org/abs/hep-th/0202140v3}{{\tt
  hep-th/0202140}}].

\bibitem{Dehghani:2004p869}
M.~H. Dehghani, ``Asymptotically (anti)-de Sitter solutions in Gauss-Bonnet
  gravity without a cosmological constant,'' Phys. Rev. D70: 064019, 2004 [arXiv: \href{http://www.arXiv.org/abs/hep-th/0405206v1}{{\tt
  hep-th/0405206}}].

\bibitem{Dehghani:2002p6114}
M.~H. Dehghani, ``Charged Rotating Black Branes in anti-de Sitter
  Einstein-Gauss-Bonnet Gravity,'' Phys. Rev. D67: 064017, 2003 [arXiv:
  \href{http://www.arXiv.org/abs/hep-th/0211191v1}{{\tt hep-th/0211191}}].

\bibitem{Dehghani:2006p2309}
M.~H. Dehghani and S.~H. Hendi, ``Taub-NUT/Bolt Black Holes in
  Gauss-Bonnet-Maxwell Gravity,'' Phys. Rev. D73: 084021, 2006 [arXiv:  \href{http://www.arXiv.org/abs/hep-th/0602069v2}{{\tt hep-th/0602069}}].

\bibitem{Charmousis:2002p6115}
C.~Charmousis and J.-F. Dufaux, ``General Gauss-Bonnet brane cosmology,'' Class. Quant. Grav. 19 (2002) 4671-4682 [arXiv:
  \href{http://www.arXiv.org/abs/hep-th/0202107v3}{{\tt hep-th/0202107}}].

\bibitem{Brihaye:2011p6116}
Y.~Brihaye, ``Charged, rotating black holes in Einstein-Gauss-Bonnet gravity,''
  arXiv:
  \href{http://www.arXiv.org/abs/1108.2779v1}{{\tt 1108.2779}}.
  
\bibitem{Myers:1988p5951}
R.~Myers and J.~Simon, ``Black-hole thermodynamics in Lovelock gravity,'' 
  Phys. Rev. D38, No.8 (1988) 2434.

\bibitem{Myers2}R. Myers, ``Superstring gravity and black holes,''  Nucl. Phys. B289 (1987) 701.

\bibitem{Cvetic:2001p6166}
M.~Cvetic, S.~Nojiri, and S.~D. Odintsov, ``Black Hole Thermodynamics and
  Negative Entropy in deSitter and Anti-de Sitter Einstein-Gauss-Bonnet
  gravity,'' Nucl. Phys. B628: 295-330, 2002 [arXiv:
  \href{http://www.arXiv.org/abs/hep-th/0112045v4}{{\tt hep-th/0112045}}].

\bibitem{Nojiri:2001p6167}
S.~Nojiri and S.~D. Odintsov, ``Anti-de Sitter Black Hole Thermodynamics in
  Higher Derivative Gravity and New Confining-Deconfining Phases in dual CFT,''
  Phys. Lett. B521: 87-95, 2001; Erratum-{\em ibid.} B542: 301, 2002 [arXiv:  \href{http://www.arXiv.org/abs/hep-th/0109122v3}{{\tt hep-th/0109122}}].

\bibitem{Cho:2002p5953}
Y.~Cho and I.~Neupane, ``Anti-de Sitter black holes, thermal phase transition,
  and holography in higher curvature gravity,'' Phys. Rev. D66: 024044, 2002.

\bibitem{Zhao:2011p693}
L.~Zhao, ``Note on a class of anisotropic Einstein metrics,'' arXiv: \href{http://www.arXiv.org/abs/1106.5027}{{\tt
  1106.5027}}]. 

\bibitem{Brill}D. R. Brill and J. Louko, ``Thermodynamics of (3+1)-dimensional black holes with toroidal or higher genus horizons'',  Phys. Rev. D56 (1997) 3600-3610 
[arXiv: \href{http://www.arXiv.org/abs/gr-qc/9705012v2}{{\tt gr-qc/9705012}}].

\bibitem{Emparan}R. Emparan, ``AdS Membranes Wrapped on Surfaces of Arbitrary Genus,'' Phys. Lett. B432 (1998) 74-82 [arXiv:
  \href{http://www.arXiv.org/abs/hep-th/9804031v2}{{\tt hep-th/9804031}}].
  
  
\bibitem{Birmingham}D. Birmingham, ``Topological Black Holes in Anti-de Sitter Space'', Class. Quant. Grav. 16 (1999) 1197-1205 [arXiv:
  \href{http://www.arXiv.org/abs/hep-th/9808032v3}{{\tt hep-th/9808032}}].

\end{thebibliography}

\providecommand{\href}[2]{#2}\begingroup\raggedright\endgroup

\end{document}